# Probing efficient n-type lanthanide dopants for $Mg_3Sb_2$ thermoelectrics

*Jiawei Zhang,* Lirong Song, and Bo Brummerstedt Iversen**

Dr. J. Zhang, Dr. L. Song, Prof. B. B. Iversen
Center for Materials Crystallography, Department of Chemistry and iNANO, Aarhus University, DK-8000 Aarhus, Denmark
E-mail: jiaweizhang@chem.au.dk, bo@chem.au.dk

**The recent discovery of n-type $Mg_3Sb_2$ thermoelectric has ignited intensive research activities on searching for potential n-type dopants for this material. Using first-principles defect calculations, here we conduct a systematic computational screening of potential efficient n-type lanthanide dopants for $Mg_3Sb_2$. In addition to La, Ce, Pr, and Tm, we find that high electron concentration ($\gtrsim 10^{20}$ cm$^{-3}$ at the growth temperature of 900 K) can be achieved by doping on the Mg sites with Nd, Gd, Ho, and Lu, which are generally more efficient than other lanthanide dopants and the anion-site dopant Te. Experimentally, we confirm Nd and Tm as effective n-type dopants for $Mg_3Sb_2$ since doping with Nd and Tm shows superior thermoelectric figure of merit $zT \gtrsim 1.3$ with higher electron concentration than doping with Te. Through codoping with Nd (Tm) and Te, simultaneous power factor improvement and thermal conductivity reduction are achieved. As a result, we obtain high $zT$ values of ~1.65 and ~1.75 at 775 K in n-type $Mg_{3.5}Nd_{0.04}Sb_{1.97}Te_{0.03}$ and $Mg_{3.5}Tm_{0.03}Sb_{1.97}Te_{0.03}$, respectively, which are among the highest values for n-type $Mg_3Sb_2$ without alloying with $Mg_3Bi_2$. This work sheds light on exploring promising n-type dopants for the design of $Mg_3Sb_2$ thermoelectrics.**

Thermoelectric (TE) materials show great promise in waste heat recovery and solid-state refrigeration applications since they can directly convert heat into electricity or *vice versa* purely by solid-state means.[1,2] The performance of TE materials is typically determined by the dimensionless figure of merit $zT = \alpha^2\sigma T/\kappa$, where $\alpha$ is the Seebeck coefficient, $\sigma$ is the electrical conductivity, $T$ is the absolute temperature, and $\kappa$ is the total thermal conductivity.

Low-cost high-performance materials are required for the widespread application of TE technology. The recently discovered n-type $Mg_3Sb_2$-based compound is one such material that shows exceptionally high TE performance with low-cost and earth-abundant chemical compositions. The excellent TE performance with $zT \approx 1.6$ at 725 K was first reported in n-type $Mg_3Sb_{1.5}Bi_{0.5}$ with Te as an effective electron dopant on the anion sites, where the good TE performance is dominantly attributed to the multiple conducting electron pockets and light conductivity effective mass.[3-5] Later, many experimental efforts have been made to improve the low-temperature performance of n-type $Mg_{3+\delta}Sb_{2-x}Bi_x$.[6-13] Theoretically, accurate electronic structure,[14,15] chemical bonding,[16-18] and phonon transport properties[19,20] have been revealed for understanding the outstanding TE properties of n-type $Mg_3Sb_2$-based TE materials. Despite many significant studies on n-type Te-doped $Mg_3Sb_2$-based compounds, the optimization of transport properties requires efficient n-type dopants for the broad tunability of carrier density. For the anion site doping, Se and S also have been reported as effective n-type dopants for $Mg_3Sb_2$,[14,21] whereas they are less efficient than Te due to the higher substitution defect formation energies. Recent defect calculations[22-25] predicted that the group-3 elements (Sc and Y) as well as several lanthanides including La, Pr, Ce, and Tm are efficient n-type cation-site dopants for $Mg_3Sb_2$ and doping with these elements on the Mg sites is able to achieve higher electron concentration than doping with chalcogens on the anion sites. The subsequent experiments[24,26-32] confirmed Sc, Y, La, Pr, and Ce as efficient n-type dopants on the Mg sites for $Mg_{3+\delta}Sb_{2-x}Bi_x$. Despite these significant theoretical and experimental efforts on exploring efficient n-type dopants, the n-type doping behavior of many other lanthanide dopants in $Mg_3Sb_2$ remains largely unexplored so far.

In this work, we perform first-principles defect calculations to explore potential n-type dopants from all lanthanides for $Mg_3Sb_2$ TEs. It is shown that, in addition to La, Ce, Pr, and Tm, several other lanthanides including Nd, Gd, Ho, and Lu are efficient n-type dopants on the cation sites for $Mg_3Sb_2$. The predicted free electron concentrations for doping with these lanthanide elements in $Mg_3Sb_2$ generally approach or exceed $10^{20}$ $cm^{-3}$ at the growth temperature of 900 K under the Mg-rich condition, which are higher than those of doping with the chalcogens. For the experimental validation, we have successfully synthesized n-type Nd-doped, Tm-doped, (Nd, Te)-codoped, and (Tm, Te)-codoped $Mg_3Sb_2$ without alloying with $Mg_3Bi_2$. In agreement with the theoretical calculation, the experimental electron concentrations of Nd-doped and Tm-doped $Mg_3Sb_2$ samples are indeed higher than those of the Te-doped ones. Among n-type Nd-doped and Tm-doped samples, optimal $zT$ values of 0.16-1.30 and 0.27-1.38 at 300-775 K are obtained in $Mg_{3.5}Nd_{0.04}Sb_2$ and $Mg_{3.5}Tm_{0.03}Sb_2$, respectively. By codoping with the appropriate amount of Nd (Tm) and Te, simultaneous power factor enhancement and thermal conductivity reduction can be achieved. As a result, high $zT$ values of ~1.65 and ~1.75 are obtained at 775 K, respectively, in $Mg_{3.5}Nd_{0.04}Sb_{1.97}Te_{0.03}$ and $Mg_{3.5}Tm_{0.03}Sb_{1.97}Te_{0.03}$, outperforming any reported n-type $Mg_3Sb_2$ samples without alloying with $Mg_3Bi_2$.[32-35]

The intrinsic point defects under different growth conditions play a crucial role in understanding the doping behavior of $Mg_3Sb_2$. Point defects in $Mg_3Sb_2$ can exist as either electron-producing donors with positive charges or electron-capturing acceptors with negative charges, which are indicated by defect plots with positive or negative slopes (see Figure S1, Supporting Information), respectively. Because of the high vapor pressure and easy oxidation of Mg, $Mg_3Sb_2$-based compounds prepared by traditional melt methods are usually Mg-deficient. Under the Mg-poor condition, the negatively-charged Mg vacancies, acting as the electron-capturing acceptors, are the dominant native defects showing low defect formation energies within the bulk gap (see Figure S1a, Supporting Information), which pin the Fermi level close to the valence band maximum (VBM) and thereby make pure $Mg_3Sb_2$ samples persistently p-type.[34,36]

In contrast, under the Mg-rich situation, the defect formation energies of Mg vacancies are notably increased and the Mg interstitials (donor defects) become the dominant native defects (see Figure S1b, Supporting Information), which means that the Fermi level can move above the middle of the band gap (mid-gap) and n-type conductivity can be achieved. However, the free electron concentration contributed by the positively-charged Mg interstitials is very low (in order of ~$10^{18}$ $cm^{-3}$)[22,23] since the compensating acceptor defects Mg vacancies can spontaneously form when the Fermi level is close to the conduction band minimum (CBM), which pin the Fermi level well below the CBM. Thus, the efficient n-type doping in $Mg_3Sb_2$ requires Mg-rich growth condition as well as the extrinsic donor defects with sufficiently low formation energies to avoid the electron compensation by the negatively-charged Mg vacancies.

To explore potential efficient n-type dopants for $Mg_3Sb_2$, we investigated the effect of n-type doping with all lanthanide elements. The formation energies of all the extrinsic defects $Ln_{Mg}$ (i.e., Mg substitution with lanthanides Ln) were calculated using density functional theory. The results as well as their comparison with the native defects and other extrinsic defects ($Te_{Sb}$, $Sc_{Mg}$, and $Y_{Mg}$) under Mg-rich conditions are shown in **Figure 1**a,b and Figure S2-S5 in the Supporting Information. There are two nonequivalent Mg atoms in $Mg_3Sb_2$, where the Mg1 atom occupies the octahedral site between the $[Mg_2Sb_2]^{2-}$ slabs while the Mg2 atom resides in the tetrahedral site within the $[Mg_2Sb_2]^{2-}$ slabs.[37] The formation energies of the extrinsic defects $Ln_{Mg1}$ are lower than those of $Ln_{Mg2}$, which indicates that the lanthanides prefer to substitute Mg1 over Mg2. This might be attributed to the Mg1 atom showing relatively weaker adjacent bonds as well as the larger void space induced by the octahedral site.[16,20] All extrinsic defects $Ln_{Mg1}$ generally show no electron compensation by the native defects within the bulk gap since they show much lower formation energies than those of the Mg vacancies and the extrinsic defect $Te_{Sb}$. In defect plots, the slope of the energy line corresponds to the charge state of the defect and the Fermi energy at which the slope changes represents the transition energy level. The substitution defects $Ln_{Mg1}$ generally have +1, 0, and -1 charge states. For n-type doping, it is favorable to form the donor defect with the charge state +1, while it is detrimental to form the defect with the charge state -1 (or 0) since it can compensate (or limit) the electron concentration.

In general, to be efficient n-type dopants, they not only should have reasonably low formation energies to ensure good solubility but also should show high transition energy levels (+1/0), (0/-1) and (+1/-1) very close to or well above the CBM so that the Fermi level is able to shift close to or above the CBM. It is clear that Eu and Yb are very unfavorable as n-type dopants since $Eu_{Mg1}$ and $Yb_{Mg1}$ show very low (or deep) (+1/0) transition levels even close to or below the VBM, which pin the Fermi level only slightly above the mid-gap (see Figure S4e,f, Supporting Information). In contrast, Y, La, Ce, and Tm are very efficient as n-type dopants for $Mg_3Sb_2$ because the extrinsic defects $Ln_{Mg1}$ (Ln = Y, La, Ce, and Tm) show high donor transition levels (+1/0) and (+1/-1) well above the CBM (see Figure 1b and Figure S2c-e and S3d, Supporting Information). Although the extrinsic defect $Pr_{Mg1}$ shows a bit low donor transition level (+1/0) below the CBM, very high transition level (0/-1) located at ~0.5 eV above the CBM as well as the low formation energy allows the Fermi energy to shift well above the CBM (see Figure 1b and Figure S2f, Supporting Information), making Pr a very efficient n-type dopant for $Mg_3Sb_2$. In addition to Y, La, Ce, Pr, and Tm, it is obvious that Nd, Gd, Lu, and Ho are strong candidates as efficient n-type dopants for $Mg_3Sb_2$ since they show low formation energies as well as the shallow donor transition level (+1/0) close to the CBM (see Figure 1a,b and Figure S3a-c, Supporting Information). These dopants are typically more efficient than other lanthanides such as Sm, Dy, Er, Pm, and Tb, which, however, show low transition levels (+1/0) and (0/-1) that pin the Fermi level below the CBM (Figure 1b and Figure S4a-d, Supporting Information).

Figure 1c and Figure S6 in the Supporting Information show the calculated free electron concentrations for the efficient n-type dopants as well as the comparison of free electron concentrations at the growth temperature of 900 K for all lanthanide dopants with those of n-type Te, Sc, and Y dopants in $Mg_3Sb_2$ under the Mg-rich condition. Consistent with the results of defect formation energies (see Figure S4, Supporting Information), Eu and Yb are largely ineffective as n-type dopants showing poor free electron concentrations. It is evident that Sm, Dy, Er, Pm, and Tb are a bit less efficient and show lower free electron concentrations than Te for n-type doping since they show a bit deeper donor levels even though they have lower formation energies. Besides La, Ce, Pr, and Tm that have been confirmed by previous calculations,[22,24,25] Nd, Gd, Lu, and Ho, as expected, are more effective as n-type dopants than Te and doping with these lanthanide elements results in high maximum achievable electron concentrations $\gtrsim 10^{20}$ cm$^{-3}$ at 900 K. Among these efficient n-type dopants, Nd and Tm show the lowest formation energies of the donor defects $Nd_{Mg1}$ (+1) and $Tm_{Mg1}$ (+1) throughout the bulk gap. The predicted maximum achievable free electron concentrations at the growth temperature of 900 K for the potential n-type dopants Nd, Gd, Lu, and Ho are respectively $9.90 \times 10^{19}$, $1.30 \times 10^{20}$, $1.27 \times 10^{20}$, and $1.17 \times 10^{20}$ cm$^{-3}$, which are slightly lower than those ($> 2 \times 10^{20}$ cm$^{-3}$) of the n-type dopants La, Ce, and Pr. Recent experiments[24,26,31] have already confirmed La, Ce, and Pr as effective n-type dopants for $Mg_3Sb_{2-x}Bi_x$, while other predicted efficient n-type dopants such as Nd, Gd, Lu, Ho, and Tm still require the experimental confirmation.

To verify the theoretical prediction, here we experimentally examine the potential of Nd and Tm as n-type dopants for $Mg_3Sb_2$. The Nd-doped, Tm-doped, (Nd, Te)-codoped, and (Tm, Te)-codoped $Mg_3Sb_2$ bulk samples were prepared by the previously-proposed spark plasma sintering method[30,38] followed by annealing in the Mg-rich condition.[11] According to our previous work,[30] the specific amount of excess Mg ($Mg_{3.5}$) is used to compensate for the evaporation loss of Mg so that the as-SPS-pressed pellets show no Mg deficiency. All as-synthesized bulk samples are dense with relative densities larger than 95% (Table S1, Supporting Information). Powder X-ray diffraction (PXRD) patterns confirm that all samples are virtually phase pure with a very small amount of MgO (see **Figure 2**a,b and Figure S7-S8, Supporting Information). The Nd-doped sample $Mg_{3.5}Nd_{0.04}Sb_2$ and all (Nd, Te)-codoped samples show a small amount of secondary phase, which can be indexed to the NdSb cubic phase (space group: $Fm\bar{3}m$). Similarly, a small amount of secondary phase TmSb is found in several $Mg_{3.5}Tm_ySb_{2-x}Te_x$ samples ($y$ = 0.03-0.04, $x$ = 0 and 0.03). For Nd-doped samples $Mg_{3.5}Nd_ySb_2$, the lattice parameters show a gradually increasing trend as the fraction $y$ increases from 0.01 to 0.03 (see Table S2, Supporting Information). This can be understood by the larger ionic radius of $Nd^{3+}$ than that of $Mg^{2+}$. On the other hand, the lattice parameters decrease when the fraction $y$ increases from 0.03 to 0.04, which is likely induced by the formation of the secondary phase NdSb. A similar trend in the lattice parameters is found in Tm-doped samples (Table S2, Supporting Information), which can be understood in the same manner.

The EDS elemental mapping of the $Mg_{3.5}Nd_{0.03}Sb_{1.97}Te_{0.03}$ sample shows that the sample is virtually homogeneous with nearly uniform distributions of the constituent elements (Figure S9, Supporting Information). The chemical composition measured by SEM-EDS reveals that the actual Mg content of the $Mg_{3.5}Nd_{0.03}Sb_{1.97}Te_{0.03}$ sample increases from $Mg_{3.05}$ to $Mg_{3.3}$ after annealing in the Mg-rich condition in an Argon atmosphere (Table S3, Supporting Information). The evolution of the PXRD patterns confirms that the amount of MgO shows nearly no change after annealing (Figure S8, Supporting Information), which indicates that the 10% excess Mg after annealing should not exist dominantly as the MgO secondary phase at the grain boundaries. The increase in cell parameters after annealing in the Mg-rich condition could be an indication of the excess Mg partially entering the interstitial sites of the structure (Table S4, Supporting Information). However, only a small proportion of the excess Mg might exist as the Mg interstitials within the structure due to the small solubility.[34] Therefore, the excess Mg should be dominantly contributed by the elemental Mg that diffuses into the grain boundaries of the sample during the Mg-rich annealing, which is confirmed by the elemental Mg peaks in the PXRD pattern of the $Mg_{3.5}Nd_{0.03}Sb_{1.97}Te_{0.03}$ sample right after the annealing (Figure S8, Supporting Information).

For the better comparison of the doping effect on the carrier concentration and mobility in n-type $Mg_3Sb_2$, several Y-doped samples $Mg_{3.5}Y_ySb_2$ ($y$ = 0.01-0.04) were also prepared using the same method as for Nd-doped samples and the corresponding experimental results are shown in Figure S10-S13 (Supporting Information). With the experimental data of the Nd-, Tm-, and Y-doped samples, we can compare the results with those of the reported Te- and Sc-doped samples prepared using a similar synthesis method.[30] It is found that n-type Nd-doped and Tm-doped $Mg_3Sb_2$ samples show slightly lower electron concentrations than those of the Y-doped samples but higher electron concentrations than those of Sc-doped and Te-doped samples (see Figure 2c), which is consistent with the theoretical calculation. The room temperature electron concentrations of n-type $Mg_{3.5}Nd_ySb_2$ samples vary from $1.46 \times 10^{19}$ ($y$ = 0.01) to $3.17 \times 10^{19}$ cm$^{-3}$ ($y$ = 0.03), while the electron concentrations vary from $7.32 \times 10^{18}$ to $2.36 \times 10^{19}$ cm$^{-3}$ in $Mg_{3.5}Tm_ySb_2$ samples. Nearly all experimental electron concentration data points of Nd- and Tm-doped samples fall exactly within the pink region shown in Figure 2d with the predicted free electron concentration at the Mg-rich and Mg-poor condition as the upper and lower bound, respectively. Further improvement of the electron concentration of Nd-doped and Tm-doped samples can be expected through reducing the band gap via alloying with $Mg_3Bi_2$.

**Figure 3** shows the temperature-dependent electron mobility of all samples. Although the Nd-doped samples have been sintered at a high temperature of 800 °C followed by annealing at 615 °C under the Mg-rich environment[11] (see Methods) to minimize the grain boundary scattering, the electron mobility of Nd-doped $Mg_3Sb_2$ samples still shows a temperature dependence of $T^{1.5}$ at low temperatures. This is very different from the Y-doped, Sc-doped,[30] or Te-doped[30] samples synthesized using a similar method, which show dominant acoustic phonon scattering at low temperatures (Figure S14, Supporting Information). The increasing temperature dependence of the mobility at low temperatures results in poor room-temperature electron mobility (~20.8-32.3 $cm^2\ V^{-1}\ s^{-1}$) of Nd-doped samples in comparison with those of Te-doped samples (~51.0-73.8 $cm^2\ V^{-1}\ s^{-1}$).[30] By codoping with Nd and Te, acoustic phonon scattering becomes dominant throughout the entire temperature range as indicated by the temperature dependence of the mobility following $T^{-p}$ ($1 \leq p \leq 1.5$). As a result, the (Nd, Te)-codoped $Mg_3Sb_2$ samples show enhanced room-temperature electron mobility values of ~38.4-54.6 $cm^2\ V^{-1}\ s^{-1}$, which, however, are still lower than those of many reported Te-doped $Mg_3Sb_2$ samples.[30]

The low carrier mobilities in Nd-doped samples in comparison with Y-, Sc-, and Te-doped samples (see Figure 3a and Figure S14 and S15a, Supporting Information) may be induced by the impurity states near the CBM introduced by the Nd doping. The low carrier mobilities at low temperatures in Nd-doped samples may be attributed to the impurity band conduction (Note S1, Supporting Information).[39] By comparing the calculated density of states (DOS) of the defect structure $Mg_{53}Nd_1Sb_{36}$ ($Mg_{2.944}Nd_{0.056}Sb_2$) with the perfect structure $Mg_{54}Sb_{36}$ ($Mg_3Sb_2$), it is found that the *f* states of the impurity atom Nd exist largely below the CBM within the bulk gap (Figure S15b, Supporting Information). Moreover, the Nd *f* states show hybridization with the electronic states (Mg 3*s* states) of the near-edge conduction bands resulting in the enhanced DOS of the CBM. The relatively localized feature of the *f* states below the CBM as well as the enhanced DOS (effective mass) of the CBM induced by the hybridization of the *f* states with the near-edge electronic states is likely the origin of low electron mobility in Nd-doped $Mg_3Sb_2$. The appearance of the *f* states below the CBM as well as their hybridization with the conduction band in Nd-doped samples can be attributed to the relatively small energy separation between the atomic orbital levels of the Nd *f* states and Mg 3*s* states. Similarly, the above mechanism may be used to explain the low electron mobilities in previously-reported La-, Pr-, and Ce-doped $Mg_3Sb_{2-x}Bi_x$ samples[24,26,31] because of the small energy separation between the atomic orbital energies of the *f* states of the dopants (La, Pr, and Ce) and 3*s* states of Mg (see Table S5, Supporting Information).

In comparison with Nd-doped samples, Tm-doped $Mg_3Sb_2$ samples generally show clearly higher electron mobilities especially at low temperatures. Room-temperature electron mobilities (~48.8-60.7 $cm^2$ $V^{-1}$ $s^{-1}$) of Tm-doped samples are comparable to those of many reported Te-doped $Mg_3Sb_2$ samples.[30] This is because that the *f* states of the Tm dopant mainly contributed to the electronic states of the valence bands (rather than within the bulk gap and conduction bands) in Tm-doped $Mg_3Sb_2$ due to the much lower atomic orbital energies of the Tm *f* states (see Figure S16 and Table S5, Supporting Information).

**Figure 4** and **Figure 5** show the TE properties of n-type Nd-doped, Tm-doped, (Nd, Te)-codoped, and (Tm, Te)-codoped $Mg_3Sb_2$ samples. The Seebeck coefficient and electrical resistivity of all samples show typical degenerate semiconductor behaviors with increasing temperature dependence at elevated temperatures. The resistivity of several Nd-doped and Tm-doped samples shows a clear decreasing trend with increasing temperature at low temperatures, which is induced by the increasing temperature dependence of the mobility. Unlike the reported Te-doped samples showing clear hysteresis,[4] the resistivity data of $Mg_{3.5}Nd_ySb_2$ and $Mg_{3.5}Tm_ySb_2$ samples show virtually no hysteresis upon thermal cycling (Figure S17 and S18, Supporting Information), indicating the enhanced stability and good repeatability. With increasing *y* from 0.01 to 0.03 in $Mg_{3.5}Nd_ySb_2$ and $Mg_{3.5}Tm_ySb_2$, the resistivity and absolute values of the Seebeck coefficient decreases as the electron concentration increases (see Figure 4a,b and Figure 5a,b). With the Nd and Te codoping, the resistivity can be effectively reduced in $Mg_{3.5}Nd_ySb_{1.97}Te_{0.03}$ mainly due to the enhanced electron mobility. As a result of the reduced resistivity and moderate Seebeck coefficient, the power factor of n-type (Nd, Te)-codoped samples exhibits a significant enhancement over the Nd-doped samples (Figure 4c). Because of lower resistivities induced by higher electron mobilities, the power factors of Tm-doped samples are generally higher than those of Nd-doped samples. With the proper Tm and Te codoping, the power factors can be further improved up to ~12.5-16.0 μW $cm^{-1}$ $K^{-2}$ within 300-725 K in $Mg_{3.5}Tm_{0.02}Sb_{1.97}Te_{0.03}$ (Figure 5c). The total thermal conductivity values of Nd-doped and Tm-doped samples are generally lower than those of the reported Te-doped[30] samples prepared using the same method (Figure 4d and Figure 5d). By codoping with the appropriate amount of Nd (Tm) and Te, the thermal conductivity is effectively reduced to 0.54 W $m^{-1}$ $K^{-1}$ (0.53 W $m^{-1}$ $K^{-1}$) at 775 K in $Mg_{3.5}Nd_{0.04}Sb_{1.97}Te_{0.03}$ ($Mg_{3.5}Tm_{0.03}Sb_{1.97}Te_{0.03}$).

As shown in **Figure 6**a, for $Mg_{3.5}Nd_ySb_2$ samples, an optimal $zT$ of 1.30 at 775 K is obtained in the sample with $y$ = 0.04, which is superior to the reported Te-doped $Mg_3Sb_2$ samples.[30,34] In comparison with Nd-doped samples, a higher $zT$ of 0.27-1.38 at 300-775 K is achieved in n-type $Mg_{3.5}Tm_{0.03}Sb_2$ (see Figure 6b), which outperforms other n-type singly-doped $Mg_3Sb_2$ samples.[30,32-35] This confirms Nd and Tm as effective n-type dopants for $Mg_3Sb_2$, which is in agreement with the theoretical prediction. Owing to the combined effect of the enhanced power factor and low thermal conductivity, the (Nd, Te)-codoped and (Tm, Te)-codoped samples generally show strongly enhanced $zT$ values in comparison with those of singly-doped samples. Peak $zT$ values of 1.65 and 1.75 at 775 K are achieved, respectively, in $Mg_{3.5}Nd_{0.04}Sb_{1.97}Te_{0.03}$ and $Mg_{3.5}Tm_{0.03}Sb_{1.97}Te_{0.03}$, which are higher than those of previously reported n-type $Mg_3Sb_2$ samples[30,32-35] without alloying with $Mg_3Bi_2$ (see Figure 6c). Considering most of the previous reports[32-35] showing $zT$ smaller than unity in n-type $Mg_3Sb_2$ without the $Mg_3Bi_2$ alloying, the high $zT$ reported in this work represents a great advance for the development of high-performance n-type $Mg_3Sb_2$ without alloying with $Mg_3Bi_2$.

In summary, we have conducted a complete computational screening of promising n-type dopants from all lanthanide elements for $Mg_3Sb_2$ thermoelectrics. In addition to La, Ce, Pr, and Tm that have been previously predicted, it is found that Nd, Gd, Lu, and Ho are strong candidates as efficient n-type cation-site dopants for $Mg_3Sb_2$ and the maximum achievable free electron concentrations for doping with these lanthanides reach $\gtrsim 10^{20}$ cm$^{-3}$ at the growth temperature of 900 K, which is higher than that of the anion-site doping with Te. For the experimental validation, we report successful n-type doping in $Mg_3Sb_2$ with Nd and Tm. In agreement with our theoretical prediction, Nd and Tm are experimentally confirmed as efficient n-type dopants for $Mg_3Sb_2$, which indeed show higher experimental electron concentration than n-type doping with Te. For n-type Nd-doped and Tm-doped $Mg_3Sb_2$ samples, we obtain high $zT$ values of 1.30 and 1.38 at 775 K, respectively. With the (Nd, Te)-codoping and (Tm, Te)-codoping in $Mg_3Sb_2$, we achieve optimal $zT$ values of 1.65 and 1.75 at 775 K, respectively, superior to any reported n-type $Mg_3Sb_2$ samples without alloying with $Mg_3Bi_2$. We can expect further improvements in TE performance through optimizing carrier concentration as well as reducing thermal conductivity via alloying with $Mg_3Bi_2$. This work provides insight into the exploration of potential effective n-type dopants for the development of $Mg_3Sb_2$ TE materials using an integrated computational and experimental approach.

**Experimental section**

*Theoretical calculations*: All calculations in this work were conducted using the projector augmented wave method[40,41] in the VASP[42] code. The defect calculations were conducted in a 3×3×2 supercell with 90 atoms using the hybrid functional HSE06[43]. A mixing parameter of 25% was used for the calculations with the HSE06 functional. The energy and Hellmann-Feynman force convergence criteria were $10^{-4}$ eV and 0.01 eV $Å^{-1}$, respectively. An energy cutoff of 400 eV was used for the plane-wave expansions. A Γ-centered 2×2×2 *k* mesh was applied for the crystal structure optimization and the total energy calculation. For the optimization of defect structures, the lattice parameters were fixed at the optimized values of the perfect supercell and all atomic positions in the defect supercell were allowed to be fully relaxed into their equilibrium positions. The symmetry was switched off during the structural relaxation. Spin polarization was included in all defect calculations. Due to the slow convergence of structural relaxations, the extrinsic defects $Sm_{Mg}$, $Dy_{Mg}$, $Tb_{Mg}$, $Er_{Mg}$ were calculated using the PBE functional[44] and a Γ-centered 4×4×4 *k* mesh with the band edge shifting correction ($\Delta E_{VBM}$ = -0.24 eV, $\Delta E_{CBM}$ = 0.18 eV) so that the band gap is corrected to be equal to the result (0.51 eV) with the HSE06 functional. The density of states of the relaxed defect supercell $Mg_{53}Nd_1Sb_{36}$ ($Mg_{53}Tm_1Sb_{36}$) with one Nd (Tm) atom substituting on the Mg1 site and the bulk supercell $Mg_{54}Sb_{36}$ shown in Figure S15b and S16 in the Supporting Information were calculated with the TB-mBJ potential[45] and a Γ-centered 6×6×6 *k* mesh.

For a point defect d with a charge state $q_d$, the formation energy can be calculated using:[46,47]

$$\Delta E_{\rm f}^{\rm d} = E_{\rm tot}^{\rm d} - E_{\rm tot}^{\rm bulk} - \sum_i n_i \mu_i + q_{\rm d}(\varepsilon_{\rm F} + E_{\rm V}) + E_{corr} \quad (1)$$

Here $E_{\rm tot}^{\rm d}$ is the total energy of the supercell with the point defect d, and $E_{\rm tot}^{\rm bulk}$ is the total energy of the perfect supercell. $n_i$ represents the number of atoms of type *i* that is added to ($n_i$ > 0) or removed from ($n_i$ < 0) the supercell when the point defect is formed, and $\mu_i$ represents the atomic chemical potentials of these species. $\varepsilon_F$ is the Fermi level, which is referenced to the valence-band maximum $E_V$ of the perfect bulk structure. $E_{corr}$ is the correction term induced by the finite size of charged supercells. Here, the image charge and potential alignment corrections for charged defects were taken into account.[48]

In $Mg_3Sb_2$, the chemical potentials of Mg and Sb must fulfill the thermodynamic stability condition: $3\Delta\mu_{Mg} + 2\Delta\mu_{Sb} = 5\Delta H_f(Mg_3Sb_2)$. To avoid the precipitations of the constituent elements, the chemical potentials should satisfy the condition: $\Delta\mu_\alpha = \mu_\alpha - \mu_\alpha^0 \leq 0$, in which $\mu_\alpha^0$ denotes the total energy per atom in the pure bulk crystal of the element $\alpha$. Under the Mg-rich condition, the chemical potentials of Q (Q = Y, Sc, La, Ce, Pr, Pm, Nd, Sm, Gd, Tb, Dy, Ho, Er, Tm, and Lu) are limited by the formation energies of the competing phases QSb ($\Delta\mu_Q + \Delta\mu_{Sb} \leq 2\Delta H_f(\mathrm{QSb})$), the chemical potentials of M (M = Eu and Yb) are constrained by $M_5Sb_3$ ($5\Delta\mu_M + 3\Delta\mu_{Sb} \leq 8\Delta H_f(M_5Sb_3)$), and the chemical potential of Te is limited by MgTe ($\Delta\mu_{Mg} + \Delta\mu_{Te} \leq 2\Delta H_f(\mathrm{MgTe})$). For the calculations of defect formation energies of $Q_{Sb}$ (Q = Nd, Gd, Ho, Lu, and Tm) under the Sb-rich condition ($\Delta\mu_{Sb}$ = 0), the chemical potentials of Q are also limited by the formation energies of QSb. Formation energies of the bulk phase and competing phases are in units of eV per atom.

Under the thermodynamic equilibrium, the Fermi energy at different growth temperatures can be obtained by self-consistently solving the charge neutrality equation:[36,46]

$$n_h - n_e + \sum_d q_d c_d = 0. \tag{2}$$

Taking the approximation of the free energy by the defect formation energy, the defect concentration can be calculated using[46,47]

$$c_d = c_0^d e^{-\Delta E_f^d / k_B T}, \tag{3}$$

where $c_0^{\mathrm{d}}$ is the concentration of defect sites for the point defect d, and $k_B$ is the Boltzmann constant. In the dilute limit, the electronic structure is assumed to be the same as that of the perfect neutral bulk cell. The numbers of holes and electrons are calculated, respectively, by

$$n_h = \int_{-\infty}^{E_V} N(\varepsilon)[1 - f(\varepsilon, \varepsilon_F)]d\varepsilon \tag{4}$$

$$n_e = \int_{E_C}^{\infty} N(\varepsilon) f(\varepsilon, \varepsilon_F) d\varepsilon \tag{5}$$

$E_C$ is the energy of the conduction band minimum. $f(\varepsilon, \varepsilon_F)$ is the Fermi-Dirac distribution. For $N(\varepsilon)$, we use the accurate density of states of the $Mg_3Sb_2$ unit cell calculated by the HSE06 functional with the tetrahedron method and a dense Γ-centered 16×16×10 *k* mesh. For solving the equilibrium Fermi level of the lanthanide (Ln) doping in $Mg_3Sb_2$ with Equation (2), we considered the extrinsic defects $Ln_{Mg1}$ and $Ln_{Mg2}$ together with all intrinsic defects with different charge states. The limitation of the DFT calculations for the lanthanides with *f*-electrons is described in Note S2, Supporting Information.

*Sample synthesis*: Stoichiometric amounts of high-purity Mg powder (99.8%, -325 mesh, Alfa Aesar), Sb powder (99.5%, -325 mesh, Chempur), Nd powder (99.9%, -200 mesh, Chempur), Tm powder (99.9%, -200 mesh, Chempur), Te powder (99.99%, -325 mesh, Chempur), and Y powder (99.9%, -200 mesh, Chempur) were weighed according to the nominal compositions of $Mg_{3+\delta}Nd_ySb_{2-x}Te_x$ ($\delta$ = 0.5, $y$ = 0.01-0.04, $x$ = 0 and 0.03), $Mg_{3+\delta}Tm_ySb_{2-x}Te_x$ ($\delta$ = 0.5, $y$ = 0.01-0.04, $x$ = 0 and 0.03), and $Mg_{3+\delta}Y_ySb_2$ ($\delta$ = 0.5, $y$ = 0.01-0.04), and mixed in a mixer (SpectroMill, Chemplex Industries, Inc.) without using ball pestles for 15 min. The excess Mg here is typically used for the purpose to compensate for the evaporation loss of Mg during the high-temperature synthesis. The mixed powders were loaded into a 1/2 inch diameter graphite die protected with the graphite paper. The above processes were conducted inside an argon-filled glove box. Afterwards, the mixed powders loaded in the graphite die were immediately sintered and consolidated in dynamic vacuum under a uniaxial pressure of 60

MPa using an SPS-515S instrument (SPS Syntex Inc., Japan). The SPS process was conducted at 673 K with a 10 min dwell and then heating to 1073 K with another 10 min dwell. Following the SPS, we conducted the annealing of the pellets in the Mg-rich environment similar to the procedure proposed by Wood *et al.*[11] The as-pressed pellets together with Mg turnings (99.9%, Chempur) were placed in a glassy carbon crucible with a lid and annealed in Ar flow at 888 K for 3 days. After annealing in the Mg-rich condition, the pellets were taken out and slightly polished before property measurements.

*Structure characterization*: The phase purity of the bulk samples was checked by the powder X-ray diffraction (PXRD) measured on a Rigaku Smartlab in the Bragg-Brentano (BB) geometry using a Cu K$\alpha_1$ source. The lattice parameters were determined by the Le Bail method[49] using the JANA2006[50] program. The microstructure, elemental distribution, and chemical composition were characterized using a scanning electron microscope (FEI Nova NanoSEM 600) equipped with an energy dispersive spectrometer (EDS).

*TE transport property measurements*: Measurements of electrical resistivity $\rho$ and Hall coefficient ($R_H$) were implemented using the *Van der Pauw* method[51] in dynamic vacuum (< $10^{-4}$ mbar) under a magnetic field up to 1.25 T.[52,53] Hall carrier concentration ($n_H$) and mobility ($\mu_H$) were evaluated respectively by $1/eR_H$ and $R_H/\rho$, where $e$ is the elementary charge. The Hall measurements of the pellets were carried out with the heating and cooling processes for 3 thermal cycles (300-725 K), in which the final stabilized cycle is used for discussion in the main text. The temperature-dependent Seebeck coefficients of the pellets were then measured with the slope method using chromel-niobium thermocouples in dynamic vacuum (< $10^{-4}$ mbar) on an in-house system,[53] whose geometry is similar to the one reported by Iwanaga *et al.*[54] The thermal diffusivity ($D$) was measured using the laser flash method (Netzsch, LFA457). The density ($d$) was estimated from the mass and volume of the pellets. Thermal conductivity was then determined using the formula $\kappa = dDC_P$. In the main text, the thermal conductivity and $zT$ were evaluated using the heat capacity ($C_P$) (see Figure

S19, Supporting Information) calculated with the polynomial equation proposed by Agne *et al*.[55] For comparison, the thermal conductivity and $zT$ estimated with the heat capacity ($C_P$) from the LFA457 setup using a Pyroceram 9606 standard sample were shown in Figure S20 and S21 in the Supporting Information. For simplicity, here the cooling curves of transport properties were adopted for discussion in the main text. The Seebeck coefficient, electrical resistivity, and thermal conductivity typically have the measurement uncertainties of 7%, 5%, and 7%, respectively.[56,57] The uncertainty for $zT$ is about 20%.

**Supporting Information**
Supporting Information is available from the Wiley Online Library or from the author.

**Acknowledgements**
J. Zhang and L. Song contributed equally to this work. This work was supported by the Villum Foundation. Affiliation with the Center for Integrated Materials Research (iMAT) at Aarhus University is gratefully acknowledged. The numerical results presented in this work were obtained at the Center for Scientific Computing, Aarhus.

Received: ((will be filled in by the editorial staff))
Revised: ((will be filled in by the editorial staff))
Published online: ((will be filled in by the editorial staff))

## References

[1] G. S. Nolas, J. Sharp, H. J. Goldsmid, *Thermoelectrics: Basic Principles and New Materials Developments*, Springer, New York, **2001**.

[2] a) G. J. Snyder, E. S. Toberer, *Nat. Mater.* **2008**, *7*, 105; b) J. He, T. M. Tritt, *Science* **2017**, *357*, eaak9997; c) G. Tan, L.-D. Zhao, M. G. Kanatzidis, *Chem. Rev.* **2016**, *116*, 12123-12149.

[3] S. H. Pedersen, *Thermoelectric properties of Zintl compounds* $Mg_3Sb_{2-x}Bi_x$., (Chemistry Project for Master's Degree, Department of Chemistry, Aarhus University, **2012**). See http://chem.au.dk/fileadmin/cmc.chem.au.dk/pictures new homepage/Pedersen S H - Thermoelectric Properties of Zintl Compounds Mg3Sb2-xBix.pdf.

[4] J. Zhang, L. Song, S. H. Pedersen, H. Yin, L. T. Hung, B. B. Iversen, *Nat. Commun.* **2017**, *8*, 13901.

[5] H. Tamaki, H. K. Sato, T. Kanno, *Adv. Mater.* **2016**, *28*, 10182.

[6] J. Shuai, J. Mao, S. Song, Q. Zhu, J. Sun, Y. Wang, R. He, J. Zhou, G. Chen, D. J. Singh, Z. Ren, *Energy Environ. Sci.* **2017**, *10*, 799.

[7] J. Mao, J. Shuai, S. W. Song, Y. X. Wu, R. Dally, J. W. Zhou, Z. H. Liu, J. F. Sun, Q. Y. Zhang, C. dela Cruz, S. Wilson, Y. Z. Pei, D. J. Singh, G. Chen, C. W. Chu, Z. F. Ren, *Proc. Natl. Acad. Sci. USA* **2017**, *114*, 10548.

[8] X. Chen, H. J. Wu, J. Cui, Y. Xiao, Y. Zhang, J. Q. He, Y. Chen, J. Cao, W. Cai, S. J. Pennycook, Z. H. Liu, L. D. Zhao, J. H. Sui, *Nano Energy* **2018**, *52*, 246.

[9] T. Kanno, H. Tamaki, H. K. Sato, S. D. Kang, S. Ohno, K. Imasato, J. J. Kuo, G. J. Snyder, Y. Miyazaki, *Appl. Phys. Lett.* **2018**, *112*, 033903.

[10] J. J. Kuo, S. D. Kang, K. Imasato, H. Tamaki, S. Ohno, T. Kanno, G. J. Snyder, *Energy Environ. Sci.* **2018**, *11*, 429.

[11] M. Wood, J. J. Kuo, K. Imasato, G. J. Snyder, *Adv. Mater.* **2019**, *31*, 1902337.

[12] K. Imasato, S. D. Kang, G. J. Snyder, *Energy Environ. Sci.* **2019**, *12*, 965.

[13] J. Mao, H. Zhu, Z. Ding, Z. Liu, G. A. Gamage, G. Chen, Z. Ren, *Science* **2019**, *365*, 495.

[14] J. Zhang, L. Song, A. Mamakhel, M. R. V. Jørgensen, B. B. Iversen, *Chem. Mater.* **2017**, *29*, 5371.

[15] J. Zhang, B. B. Iversen, *J. Appl. Phys.* **2019**, *126*, 085104.

[16] J. Zhang, L. Song, M. Sist, K. Tolborg, B. B. Iversen, *Nat. Commun.* **2018**, *9*, 4716.

[17] C. Zheng, R. Hoffmann, R. Nesper, H. G. Von Schnering, *J. Am. Chem. Soc.* **1986**, *108*, 1876.

[18] X. Sun, X. Li, J. Yang, J. Xi, R. Nelson, C. Ertural, R. Dronskowski, W. Liu, G. J. Snyder, D. J. Singh, W. Zhang, *J. Comput. Chem.* **2019**, *40*, 1693.

[19] W. Peng, G. Petretto, G. M. Rignanese, G. Hautier, A. Zevalkink, *Joule* **2018**, *2*, 1879.

[20] J. Zhang, L. Song, B. B. Iversen, *npj Comput. Mater.* **2019**, *5*, 76.

[21] J. Zhang, L. Song, K. A. Borup, M. R. V. Jørgensen, B. B. Iversen, *Adv. Energy Mater.* **2018**, *8*, 1702776.

[22] P. Gorai, B. R. Ortiz, E. S. Toberer, V. Stevanovic, *J. Mater. Chem. A* **2018**, *6*, 13806.

[23] P. Gorai, E. S. Toberer, V. Stevanović, *J. Appl. Phys.* **2019**, *125*, 025105.

[24] J. Li, F. Jia, S. Zhang, S. Zheng, B. Wang, L. Chen, G. Lu, L. Wu, *J. Mater. Chem. A* **2019**, *7*, 19316.

[25] J. Li, S. Zhang, S. Zheng, Z. Zhang, B. Wang, L. Chen, G. Lu, *J. Phys. Chem. C* **2019**, *123*, 20781.

[26] K. Imasato, M. Wood, J. J. Kuo, G. J. Snyder, *J. Mater. Chem. A* **2018**, *6*, 19941.

[27] S. W. Song, J. Mao, M. Bordelon, R. He, Y. M. Wang, J. Shuai, J. Y. Sun, X. B. Lei, Z. S. Ren, S. Chen, S. Wilson, K. Nielsch, Q. Y. Zhang, Z. F. Ren, *Mater. Today Phys.* **2019**, *8*, 25.

[28] X. Shi, T. Zhao, X. Zhang, C. Sun, Z. Chen, S. Lin, W. Li, H. Gu, Y. Pei, *Adv. Mater.* **2019**, *31*, 1903387.

[29] X. Shi, C. Sun, X. Zhang, Z. Chen, S. Lin, W. Li, Y. Pei, *Chem. Mater.* **2019**, *31*, 8987.

[30] J. Zhang, L. Song, B. B. Iversen, *Angew. Chem. Int. Ed.* **2020**, *59*, 4278; *Angew. Chem.* **2020**, *132*, 4308.

[31] F. Zhang, C. Chen, S. Li, L. Yin, B. Yu, J. Sui, F. Cao, X. Liu, Z. Ren, Q. Zhang, *Adv. Electron. Mater.* **2020**, *6*, 1901391.

[32] Y. Wang, X. Zhang, Y. Liu, Y. Wang, H. Liu, J. Zhang, *J. Materiomics* **2020**, *6*, 216.

[33] S. Kim, C. Kim, Y.-K. Hong, T. Onimaru, K. Suekuni, T. Takabatake, M.-H. Jung, *J. Mater. Chem. A* **2014**, *2*, 12311.

[34] S. Ohno, K. Imasato, S. Anand, H. Tamaki, S. D. Kang, P. Gorai, H. K. Sato, E. S. Toberer, T. Kanno, G. J. Snyder, *Joule* **2018**, *2*, 141.

[35] Y. Wang, X. Zhang, Y. Wang, H. Liu, J. Zhang, *Phys. Status Solidi (a)* **2019**, *216*, 1800811.

[36] L. Bjerg, G. K. H. Madsen, B. B. Iversen, *Chem. Mater.* **2012**, *24*, 2111.

[37] E. Zintl, E. Husemann, *Z. Phys. Chem.* **1933**, *21B*, 138.

[38] L. Song, J. Zhang, B. B. Iversen, *J. Mater. Chem. A* **2017**, *5*, 4932.

[39] T. Matsubara, Y. Toyozawa, *Prog. Theor. Phys.***1961**, *26*, 739.

[40] P. E. Blöchl, *Phys. Rev. B* **1994**, *50*, 17953.

[41] G. Kresse, D. Joubert, *Phys. Rev. B* **1999**, *59*, 1758.

[42] G. Kresse, J. Furthmüller, *Phys. Rev. B* **1996**, *54*, 11169.

[43] J. Paier, M. Marsman, K. Hummer, G. Kresse, I. C. Gerber, J. G. Angyan, *J. Chem. Phys.* **2006**, *124*, 154709.

[44] J. P. Perdew, K. Burke, M. Ernzerhof, *Phys. Rev. Lett.* **1996**, *77*, 3865.

[45] F. Tran, P. Blaha, *Phys. Rev. Lett.* **2009**, *102*, 226401.

[46] S. B. Zhang, J. E. Northrup, *Phys. Rev. Lett.* **1991**, *67*, 2339.

[47] C. G. V. d. Walle, J. Neugebauer, *J. Appl. Phys.* **2004**, *95*, 3851.

[48] C. Freysoldt, J. Neugebauer, C. G. Van de Walle, *Phys. Rev. Lett.* **2009**, *102*, 016402.

[49] A. Le Bail, *Powder Diffr.* **2012**, *20,* 316.

[50] V. Petricek, M. Dusek, L. Palatinus, *Z. Kristallogr.* **2014**, *229,* 345.

[51] L. J. v. d. Pauw, *Philips Res. Rep.* **1958**, *13,* 1.

[52] K. A. Borup, E. S. Toberer, L. D. Zoltan, G. Nakatsukasa, M. Errico, J.-P. Fleurial, B. B. Iversen, G. J. Snyder, *Rev. Sci. Instrum.* **2012**, *83*, 123902.

[53] K. A. Borup, J. de Boor, H. Wang, F. Drymiotis, F. Gascoin, X. Shi, L. Chen, M. I. Fedorov, E. Muller, B. B. Iversen, G. J. Snyder, *Energy Environ. Sci.* **2015**, *8,* 423.

[54] S. Iwanaga, E. S. Toberer, A. LaLonde, G. J. Snyder, *Rev. Sci. Instrum.* **2011**, *82,* 063905.

[55] M. T. Agne, K. Imasato, S. Anand, K. Lee, S. K. Bux, A. Zevalkink, A. J. E. Rettie, D. Y. Chung, M. G. Kanatzidis, G. J. Snyder, *Mater. Today Phys.* **2018**, *6,* 83.

[56] H. Wang, W. D. Porter, H. Böttner, J. König, L. Chen, S. Bai, T. M. Tritt, A. Mayolet, J. Senawiratne, C. Smith, F. Harris, P. Gilbert, J. Sharp, J. Lo, H. Kleinke, L. Kiss, *J. Electron. Mater.* **2013**, *42*, 1073.

[57] H. Wang, W. D. Porter, H. Böttner, J. König, L. Chen, S. Bai, T. M. Tritt, A. Mayolet, J. Senawiratne, C. Smith, F. Harris, P. Gilbert, J. W. Sharp, J. Lo, H. Kleinke, L. Kiss, *J. Electron. Mater.* **2013**, *42,* 654.

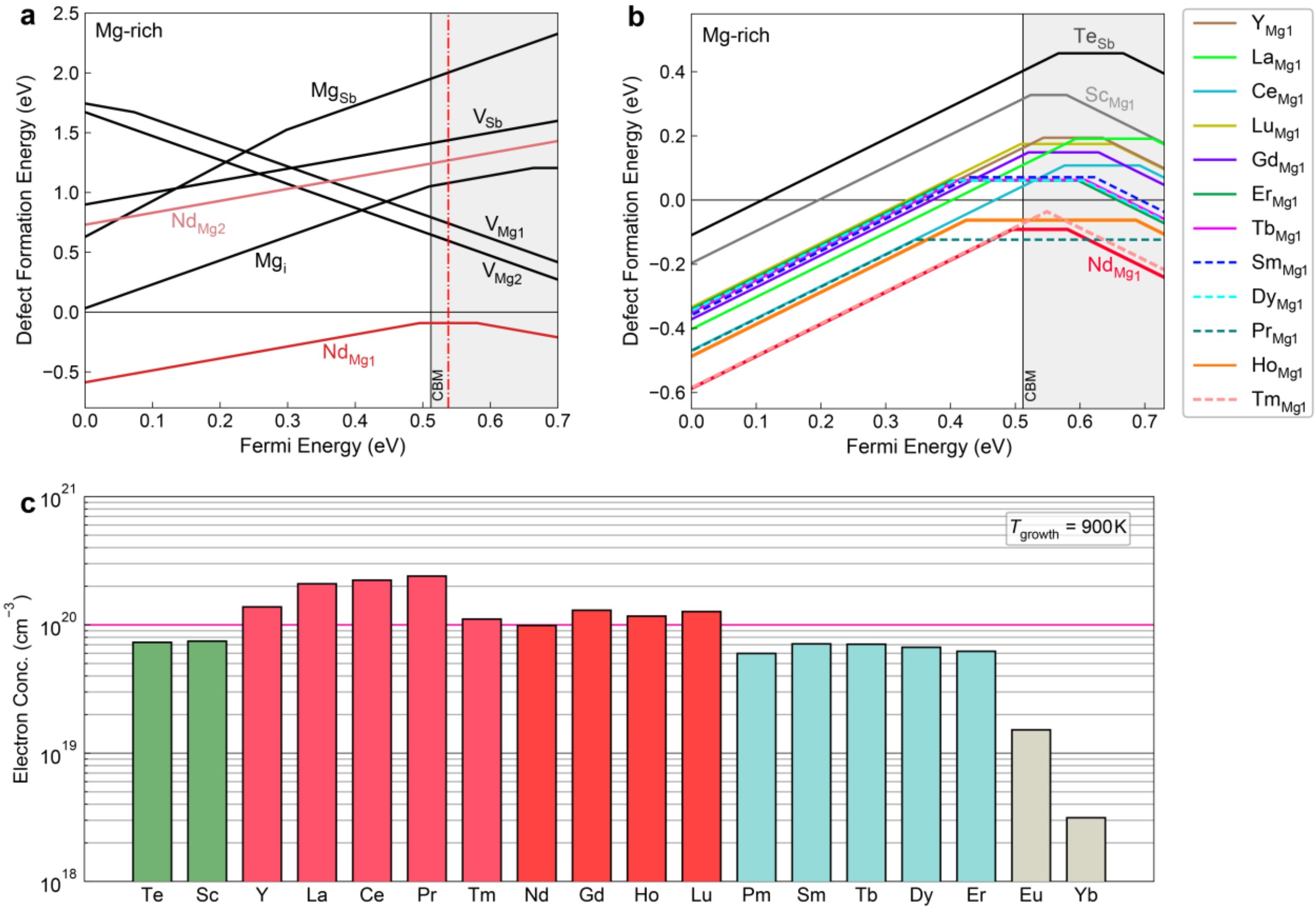

**Figure 1.** Computational screening of efficient n-type dopants from lanthanides for $Mg_3Sb_2$. a) Defect formation energies of the Nd doping on the Mg sites as well as the native defects under the Mg-rich condition. The equilibrium Fermi level at the growth temperature of 900 K is marked with the red dash-dotted line. b) Defect formation energies of the lanthanide substitution on the Mg1 site located at 1a (0, 0, 0) (denoted as $Ln_{Mg1}$) in comparison with those of the extrinsic defects $Te_{Mg1}$, $Sc_{Mg1}$, and $Y_{Mg1}$ in $Mg_3Sb_2$ under the Mg-rich condition. c) Theoretical free electron concentrations at the growth temperature of 900 K for $Mg_3Sb_2$ with 15 different n-type lanthanide dopants. The calculated electron concentrations of the Te, Sc, and Y doping for $Mg_3Sb_2$ are used for comparison.

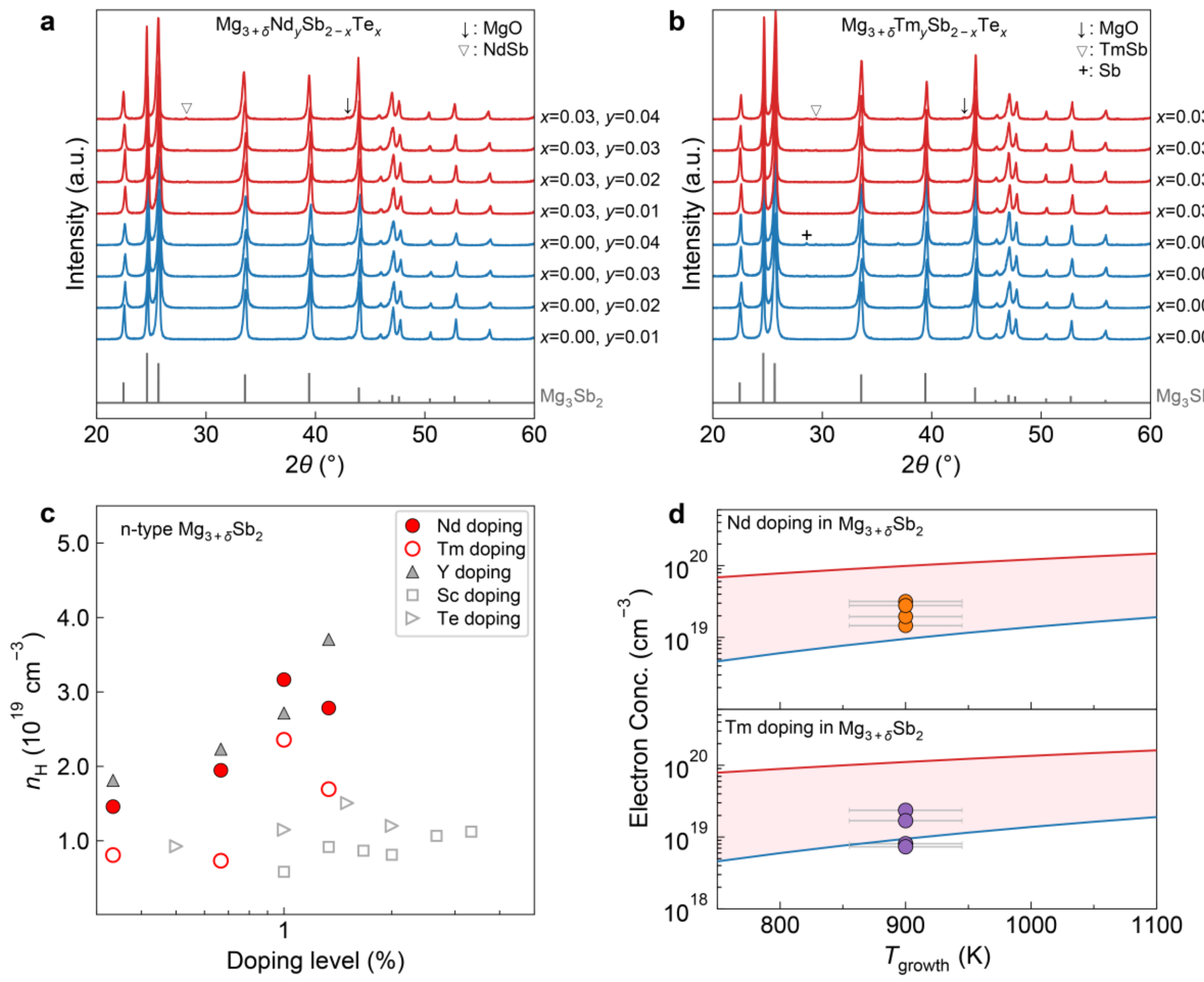

**Figure 2.** a,b) PXRD patterns of the bulk samples with the nominal compositions (a) $Mg_{3+\delta}Nd_ySb_{2-x}Te_x$ and (b) $Mg_{3+\delta}Tm_ySb_{2-x}Te_x$ ($\delta = 0.5$, $x = 0$ and 0.03, $y = 0.01$-0.04) measured before property measurements. c) Experimental Hall carrier concentration as a function of the doping level for the n-type doping with Nd and Tm comparing with doping with other elements (Y, Sc,[30] and Te[30]) in $Mg_{3+\delta}Sb_2$ without the $Mg_3Bi_2$ alloying. The data of Nd-, Tm-, and Y-doped samples are from this work. d) Predicted free electron concentration as a function of the growth temperature of the n-type doping with Nd and Tm in $Mg_3Sb_2$ under the Mg-rich and Mg-poor conditions. The orange and purple solid points represent the experimental data of n-type Nd-doped and Tm-doped $Mg_{3+\delta}Sb_2$, respectively. The growth temperature of the experimental data is set at 900 K with an uncertainty of 5%.

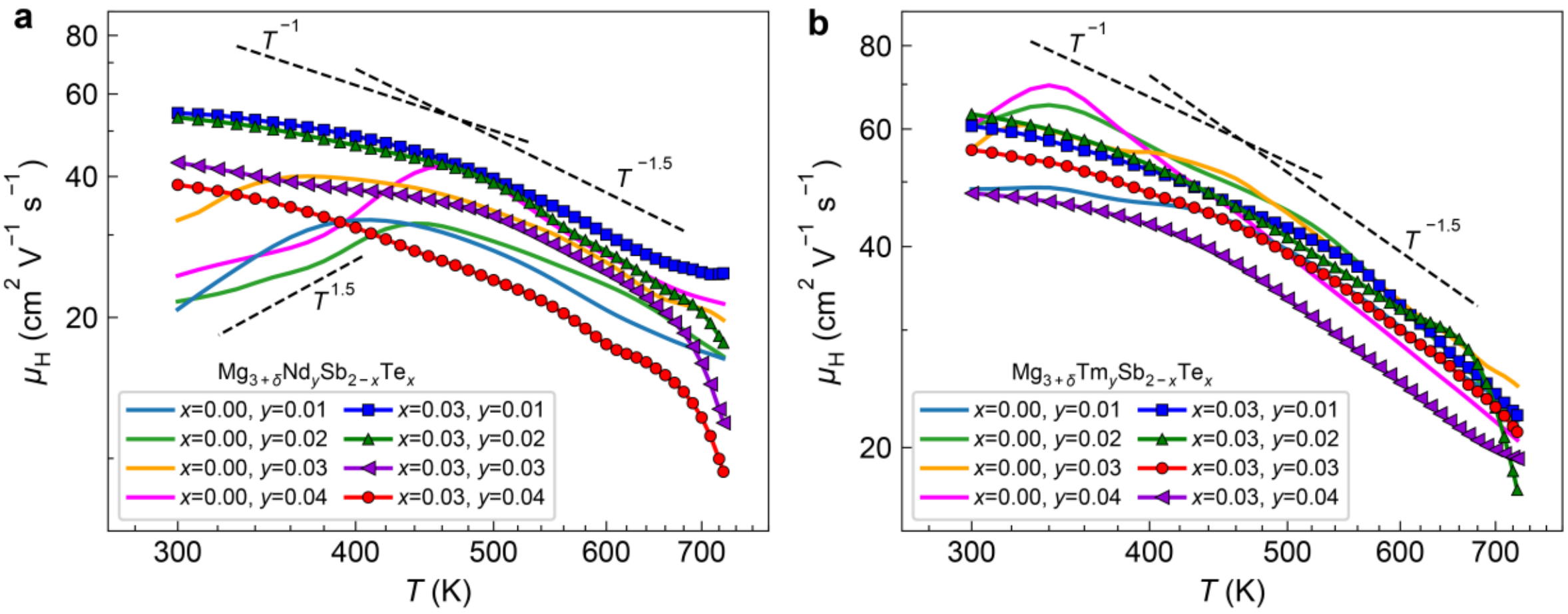

**Figure 3.** a,b) Temperature dependence of the Hall mobility of n-type (a) $Mg_{3+\delta}Nd_ySb_{2-x}Te_x$ and (b) $Mg_{3+\delta}Tm_ySb_{2-x}Te_x$ ($\delta = 0.5$).

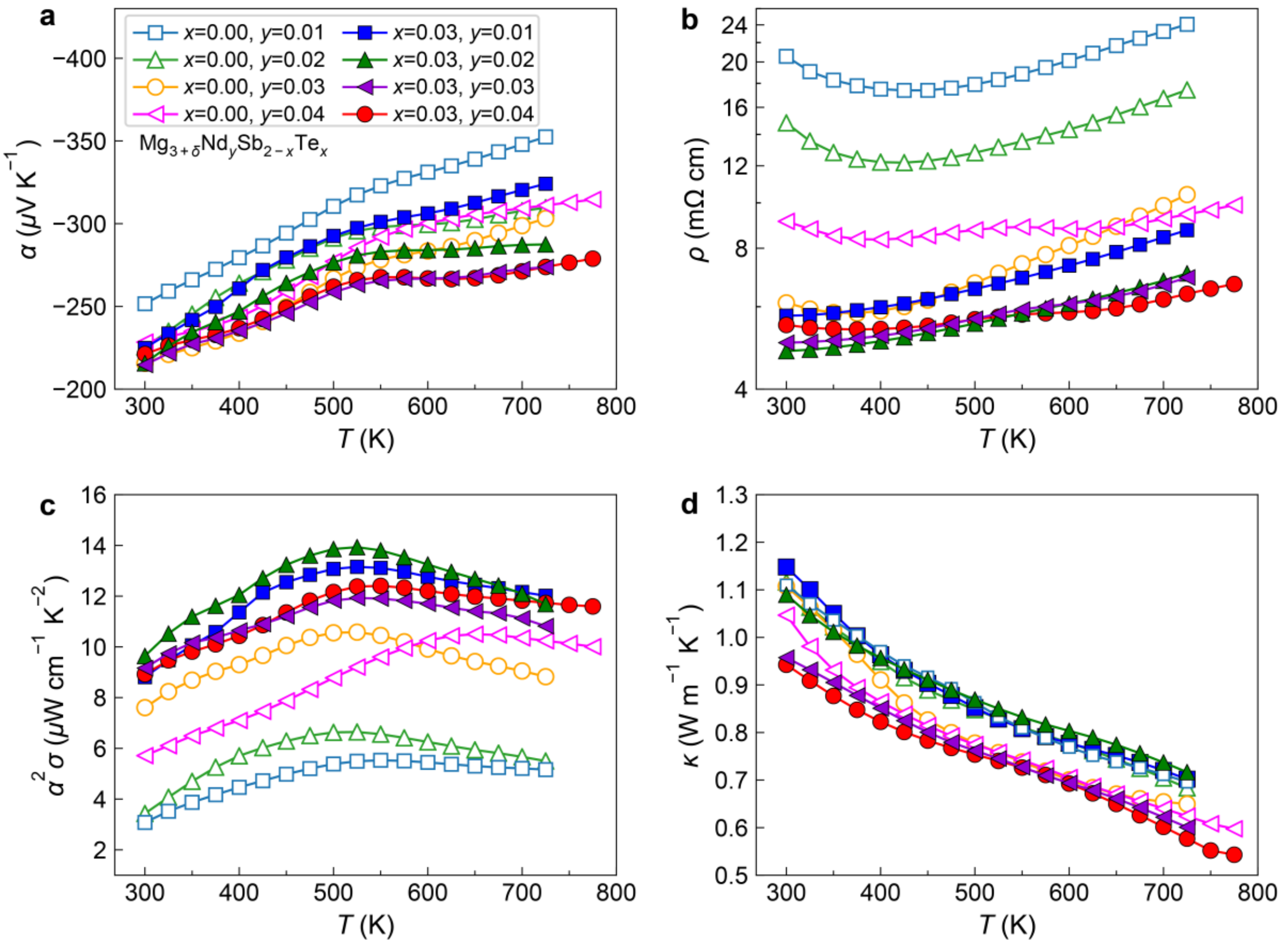

**Figure 4.** a) Temperature-dependent Seebeck coefficient, b) electrical resistivity, c) power factor, and d) total thermal conductivity of n-type $Mg_{3+\delta}Nd_ySb_{2-x}Te_x$ ($\delta$ = 0.5).

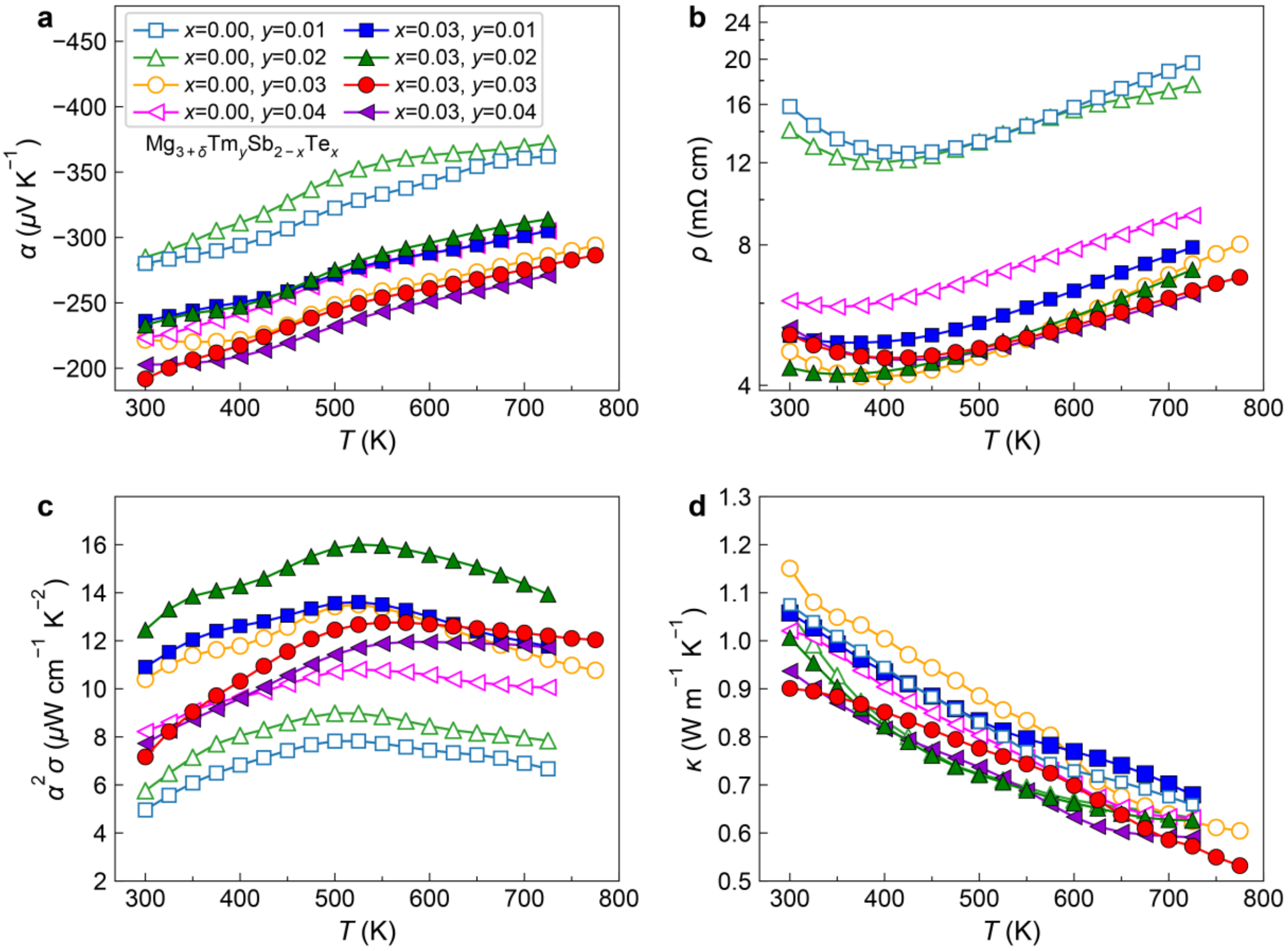

**Figure 5.** a) Temperature-dependent Seebeck coefficient, b) electrical resistivity, c) power factor, and d) total thermal conductivity of n-type $Mg_{3+\delta}Tm_ySb_{2-x}Te_x$ ($\delta = 0.5$).

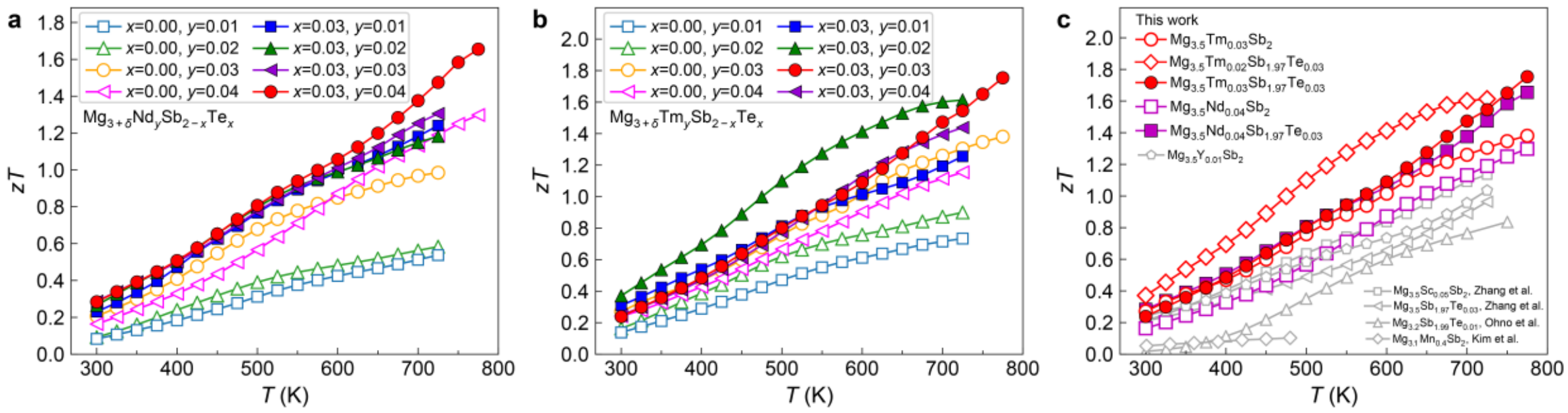

**Figure 6.** a,b) Temperature dependence of the figure of merit $zT$ of n-type (a) $Mg_{3+\delta}Nd_ySb_{2-x}Te_x$ and (b) $Mg_{3+\delta}Tm_ySb_{2-x}Te_x$ ($\delta$ = 0.5). c) $zT$ values of n-type $Mg_{3.5}Nd_{0.04}Sb_2$, $Mg_{3.5}Nd_{0.04}Sb_{1.97}Te_{0.03}$, $Mg_{3.5}Tm_{0.03}Sb_2$, $Mg_{3.5}Tm_{0.02}Sb_{1.97}Te_{0.03}$, and $Mg_{3.5}Tm_{0.03}Sb_{1.97}Te_{0.03}$ in comparison with those of the n-type Y-doped (this work), reported Te-doped,[30,34] Sc-doped,[30] and Mn-doped[33] $Mg_{3+\delta}Sb_2$ without alloying with $Mg_3Bi_2$.